\documentclass[final,notitlepage,letterpaper,11pt]{article}
\usepackage{makeidx}
\usepackage{amssymb}
\usepackage{amsmath}
\usepackage[dvips]{graphicx}
\usepackage{epsfig}
\usepackage[dvips]{geometry}
\usepackage{setspace}
\usepackage{endnotes}
\usepackage{rotating}
\usepackage{booktabs}
\usepackage{graphicx}
\usepackage{caption}
\usepackage{siunitx}
\usepackage{float}
\usepackage{changepage}

\begin{document}

\begin{onehalfspace}

\title{Do Governments React to Public Debt Accumulation?\\
A Cross-Country Analysis\thanks{%
We would like to thank Noemi Giampaoli, Giovanni Piersanti and Michele Postigliola for useful comments and discussions.
This research has received funding from the project \textquotedblleft Public
and Corporate Debt, Monetary Policy, and Macroeconomic
Stability\textquotedblright---Project Code 2022TJWFWJ\_002, CUP
I53D23002790006---funded under the National Recovery and Resilience Plan
(NRRP), Mission 4 \textquotedblleft Education and Research\textquotedblright , Component 2, Investment 1.1, \textquotedblleft Fund
for the National Research Programme and Projects of Significant National
Interest (PRIN)\textquotedblright---Call for tender No. 104 of 02/02/2022 and Concession
Decree No. 967 of 30/06/2023 of Italian Ministry of University and Research
funded by the European Union---NextGenerationEU.}}
\author{Paolo Canofari\thanks{%
Department of Economics and Social Sciences,
Polytechnic University of Marche, P.le Martelli 8, 60100 Ancona, Italy.
E-mail: p.canofari@univpm.it.} \and Alessandro Piergallini\thanks{%
Corresponding author: Department of Economics and Finance, Tor Vergata University of Rome, Via
Columbia 2, 00133 Rome, Italy. E-mail: alessandro.piergallini@uniroma2.it.}%
\vspace{0in} \and Marco Tedeschi\thanks{%
Department of Economics and Social Sciences, Polytechnic University of
Marche, P.le Martelli 8, 60100 Ancona, Italy. E-mail: m.tedeschi@univpm.it.}}
\date{July 16, 2025}
\maketitle

\vspace{-0.25in}

\end{onehalfspace}

\begin{doublespace}

\begin{abstract}

\begin{adjustwidth}{0.05in}{0.05in}

\noindent Do governments adjust budgetary policy to rising public debt, precluding fiscal unsustainability? Using budget data for $52$ industrial and emerging economies since 1990, we apply panel methods accounting for cross-sectional dependence and heterogeneous fiscal conduct. We find that a primary-balance rule with tax-smoothing motives and responsiveness to debt has robust explanatory power in describing fiscal behavior. Controlling for temporary output, temporary spending, and the current account balance, a $10$-percentage-point increase in the debt-to-GDP ratio raises the long-run primary surplus-to-GDP ratio by 0.5 percentage points on average. Corrective adjustments hold across high- and low-debt countries and across industrial and emerging economies. Our results imply many governments pursue Ricardian policy designs, avoiding Ponzi-type financing.\bigskip



\end{adjustwidth}

\end{abstract}

\newpage

\section{Introduction}

Do governments react to the accumulation of public debt, ruling out an
unsustainable course of public finances? As debt has massively increased to
counter global financial and pandemic crises, this is a central question for
public policy analysis. In this paper, we address the issue by examining
budget data in a balanced panel of 52 industrial and emerging market
economies since 1990. We infer the scope for corrective budgetary measures
should public debt embark on potential unsustainable trajectories by
employing econometric methods that account for cross-sectional dependence
and heterogeneous fiscal policy conduct. We show that a primary-balance
feedback policy rule incorporating tax-smoothing objectives and explicitly
responding to changes in outstanding debt has robust explanatory power in describing the behavior of
fiscal policymakers across countries. Specifically, we find that controlling for
temporary output, temporary spending, and the current account balance,
fiscal retrenchment exhibits a long-run upward adjustment in the primary
surplus-to-GDP ratio---by means of reducing non-interest outlays or raising
revenues---by $0.5$ percentage points on average in response to an increase
in the debt-to-GDP ratio by $10$ percentage points.

Importantly, unlike pre-2008 Global Financial Crisis studies, we show that
the conditional response of primary surpluses to debt remains significantly
positive when splitting the panel both into high- and low-debt countries and
into industrial and emerging countries. Even though the long-run budgetary
adjustment is found to be $28$ percent\ lower for the high-debt group
compared to the low-debt group and $52$\ percent lower for emerging
countries compared to industrial countries, the detected stance of
government policy is sufficiently consistent with the theoretical fiscal
requirements for public solvency prevailing in a stochastic economic environment. The
empirical results presented in this paper reveal that a large number of
governments in advanced and emerging economies do not engage in Ponzi's
games and satisfy the intertemporal budget constraint via a Ricardian\
fiscal policy design.

Our study is connected to three strands of literature. First, in our
empirical analysis we adopt an approach \`{a} la Bohn (1998, 2008) to test
the sustainability of government finance, based upon estimating a fiscal
policy reaction function and detecting whether or not the primary
surplus-to-GDP ratio is an increasing function of the debt-to-GDP ratio.\footnote{%
See D'Erasmo, Mendoza and Zhang (2016) and Canofari, Marini and Piergallini
(2020) for comprehensive reviews on the empirical strategies commonly
adopted to assess public solvency.} The basic rationale---first noted by
McCallum (1984) and then developed by Bohn (1995, 1998) a stochastic
setting---for why sustainability applies under a budget surplus stance
positively reacting to the debt-accumulation process is that public debt
turns out to grow at a lower rate relative to a Ponzi's scheme, so that its
expected present discounted value converges to zero. This implies satisfying
the private lenders' transversality condition that typically features
dynamic general equilibrium environments with forward-looking optimizing
agents (see also Benhabib, Schmitt-Groh\'{e} and Uribe, 2001, and Canzoneri,
Cumby and Diba, 2001, 2011). Whereas Bohn (1998, 2008) concentrates on the
U.S. budgetary policy, we provide cross-country panel evidence using methods
that control for both unobserved global shocks simultaneously affecting
international fiscal records and for heterogeneous stances of government
policy. A central advantage of the sustainability test we employ in our
analysis is that it does not require assumptions---which can easily be a
source of marked disagreement---about country-specific nominal or real
interest rates on government bonds, and whether or not they are above or
below country-specific nominal or real growth rates.\footnote{%
In addition, Bohn (2007) demonstrates that the empirical approaches based on
unit root tests and cointegration analyses to evaluate whether the government's
intertemporal budget constraint is satisfied are incapable of rejecting
the sustainability hypothesis.}

Second, in estimating whether primary balances can be characterized as an
increasing function of debt, we explicitly control for temporary gaps in
output and government spending, consistently with tax smoothing theory of
optimal taxation (Barro, 1979, 1986; Bohn, 1998). Indeed, when government
spending is temporarily high, for example because of wars or discretionary,
stabilization-oriented fiscal policies, or the level of output is
temporarily low, for example because of recessions, timely increases in tax
rates needed to guarantee a balanced budget would cause unnecessary economic
distortions, by influencing agents' choices for optimal time paths of labor,
production, consumption, and investment. To avoid welfare-decreasing
distortions, it turns out to be optimal for fiscal authorities to
\textquotedblleft smooth\textquotedblright \ efficiently taxes and let the
primary surplus-to-GDP ratio decline and, consequently, the debt-to-GDP ratio
raise.

Third, the panel estimation methods we employ and the quantitative results
we obtain differ from those prevailing in the existing literature. Using
standard panel techniques on pre-2008 Global Financial Crisis data, Mendoza
and Ostry (2008) find that emerging economies exhibit a stronger response of
primary balances to debt with respect to industrial economies.\footnote{%
See~Camarero, Carrion-i-Silvestre and Tamarit (2015) for an analysis entirely focused
on OECD countries.} However, the overall positive feedback behavior of
fiscal authorities vanishes for the high-debt group of countries. Our paper
differs in three important dimensions. From a methodological perspective, we
depart from Mendoza and Ostry by moving beyond standard panel models and
employing the dynamic common correlated effects estimator developed by Chudik and Pesaran (2015).
One key advantage of our adopted empirical approach is that it
accounts for both cross-sectional dependence---which arises from global
shocks that affect all countries simultaneously---and heterogeneous fiscal
policy reaction functions, thus allowing us to evaluate the issue of public
solvency more robustly. Secondly, a novel and central result that emerges
from our analysis is that, although we find significant evidence that the
2008 Global Financial Crisis has acted as a permanent negative shock on the
level of primary balances in high-debt countries, it has not significantly
altered the stance of budgetary policy in terms of the degree of
responsiveness of primary balances to debt---which is the criterion that
matters for the long-run sustainability of public finances. Finally, from a
quantitative perspective, our estimated conditional permanent adjustment of
primary surpluses to debt is 85 percent higher than that found by Mendoza
and Ostry for the overall panel---with the emerging country group reacting
52 percent lower than the industrial country group. Importantly, we document
that for the high-debt group, the estimated fiscal feedback response to debt
does not fall to a value insignificantly different from zero, as in Mendoza
and Ostry, but continues to be significantly positive. Therefore, our
results indicate that several fiscal policymakers even in highly indebted
economies take responsible corrective actions to preserve fiscal
sustainability.

The remainder of the paper is organized in four sections. In Section 2, we
set forth the theoretical requirements for fiscal sustainability and present
our empirical model specification. In Section 3, we describe the data
employed in our cross-country analysis. In Section 4, we present and discuss
the quantitative results. We close the paper in Section 5 by providing
concluding remarks.

\section{Requirements for Fiscal Sustainability and Model Specification}

Our analysis begins with the law of motion for interest-bearing public debt
derived from the government's flow budget constraint and given by%
\begin{equation*}
\frac{B_{t}}{1+r_{t}}=B_{t-1}-S_{t},
\end{equation*}%
where $B_{t}$ denotes the stock of government bonds at the end of period $t$
carried over into period $t+1$, $1+r_{t}$ the gross interest rate factor,
and $S_{t}$ the primary surplus---that is, revenues minus non-interest
expenditures. Dividing both sides by aggregate output (empirically, the
gross domestic product), $Y_{t}$, yields the law of motion for the debt-to-GDP
ratio:%
\begin{equation}
b_{t}=\frac{1+r_{t}}{1+\gamma _{t}}\left( b_{t-1}-s_{t}\right) ,  \label{B/Y}
\end{equation}%
where $b_{t}=B_{t}/Y_{t}$, $s_{t}=S_{t}/Y_{t}$, and $1+\gamma
_{t}=Y_{t}/Y_{t-1}$. Both the gross return on government bonds and the gross
growth of output can be measured either in real or nominal terms, for
inflation cancels out in the ratio. Iterating equation (\ref{B/Y}) $n$
periods forward, debt dynamics can be expressed as%
\begin{equation}
b_{t+n}=\left[ \prod_{k=0}^{n}\left( \frac{1+r_{t+k}}{1+\gamma _{t+k}}%
\right) \right] b_{t-1}-\sum_{j=0}^{n}\left[ \prod_{k=j}^{n}\left( \frac{%
1+r_{t+k}}{1+\gamma _{t+k}}\right) \right] s_{t+j}.  \label{B-dynamics}
\end{equation}

Now, from a macroeconomic perspective, the sustainability of public finance
is a general equilibrium issue, based on optimizing individual behavior, in
the sense that the government's ability to borrow is constrained by the
private agents' willingness to lend. To see this point clearly, assume
complete financial markets and infinitely-lived, forward-looking optimizing
individuals. Then, fiscal sustainability must imply the respect of the
private agents' transversality condition given by%
\begin{equation}
\lim_{n\rightarrow \infty }E_{t}\left \{ Q_{t,t+1+n}B_{t+n}\right \} =0,
\label{TVC}
\end{equation}%
where $Q_{t,t+n}$ is the pricing kernel to value at the time $t$ contingency
claims on period $t+n$. Combining the Euler equations characterizing agents'
intertemporal optimality conditions on consumption-saving decisions,%
\begin{equation*}
E_{t}\left \{ Q_{t,t+1+n}\prod \limits_{k=0}^{n}\left( 1+r_{t+k}\right)
\right \} =1\text{ \ \ }\forall \left( t,n\right) ,
\end{equation*}%
with equation (\ref{B-dynamics}) and applying the lenders' transversality
condition (\ref{TVC}) yields the following government intertemporal budget
constraint:%
\begin{equation*}
b_{t-1}=\sum \limits_{n=0}^{\infty }E_{t}\left \{ Q_{t,t+n}\prod
\limits_{k=0}^{n-1}\left( 1+\gamma _{t+k}\right) s_{t+n}\right \} .
\end{equation*}

Suppose next that government's behavior is described by a class of fiscal
policy rules of the form%
\begin{equation}
s_{t}=\phi s_{t-1}+\rho b_{t-1}+\mu _{t},  \label{fpr}
\end{equation}%
where $0<\phi <1$ is a parameter capturing the inertial policy conduct, $%
\rho >0$ is a parameter measuring the \textquotedblleft
strength\textquotedblright \ of the period-by-period primary surplus
response to debt, and $\mu _{t}$ is a bounded set of other determinants of
the primary balance. We restrict attention to the empirically plausible case
in which the long-run surplus adjustment to debt is lower than unity, that
is, $\rho /\left( 1-\phi \right) <1$. Let $L$ denote the lag operator
obeying $L^{h}x_{t}=x_{t-h}$ for any generic variable $x_{t}$. Substituting
the policy function (\ref{fpr}) into the flow budget constraint (\ref{B/Y}),
iterating $n$ periods forward, multiplying by $Q_{t,t+1+n}\prod%
\limits_{k=0}^{n}\left( 1+\gamma _{t+k}\right) $, and taking expectations,
one obtains\footnote{%
Consistently to Bohn (1998) for the $\mu _{t}$-term to be asymptotically
irrelevant, one must assume that the present value of output is finite.}%
\begin{equation*}
E_{t}\left \{ Q_{t,t+n+1}\prod \limits_{k=0}^{n}\left( 1+\gamma
_{t+k}\right) b_{t+n}\right \} =\frac{E_{t}\left \{ Q_{t,t+1+n}B_{t+n}\right
\} }{Y_{t-1}}\approx \left( 1-\frac{\rho }{1-\phi L}\right) ^{n+1}b_{t-1},
\end{equation*}%
which for any small $\rho $-value tends to zero as $n\rightarrow \infty $,
hence satisfying the lenders' transversality condition (\ref{TVC}) and
making debt sustainable. Intuitively, this is because, under the policy
function (\ref{fpr}), debt growth turns out to be permanently reduced by a
factor $1-\rho /\left( 1-\phi \right) <1$ relative to a Ponzi's scheme.

According to tax smoothing theory of optimal taxation (Barro, 1979, 1986;
Bohn, 1998), the non-debt determinants of the primary surplus $\mu _{t}$
should include measures of output and spending temporary gaps from their
respective trends, since there may be a policy scope by fiscal authorities
for \textquotedblleft smoothing\textquotedblright \ efficiently taxes in
times of recessions and/or military wars, for instance. For temporary
declines in income---and thus in the tax base---and temporary increases in
government expenditure bring about higher-than-normal budget deficits,
\textit{per se} leading to an optimal accumulation of debt in order to
minimize tax distortions on private agents' choices. In addition, since our
dataset includes many open emerging market economies, the vector $\mu _{t}$
should also include the current account balance. For external imbalances can
influence fiscal policy decisions, either by constraining borrowing capacity
or by signaling vulnerabilities that affect fiscal sustainability.

In a cross-country analysis, however, it is further essential to take into
account two key issues that typically arise in panel data modeling, known as
\textquotedblleft slope heterogeneity\textquotedblright \ (Pesaran and Yamagata, 2008; Blomquist and Westerlund, 2013) and
\textquotedblleft cross-sectional dependence\textquotedblright \ (Pesaran, 2004; Fan, Liao and Yao, 2015). Slope
heterogeneity concerns the heterogeneous responses of different
cross-sectional units and naturally arises our cross-country sustainability
analysis, due to potentially distinct policy functions adopted by
governments of different countries. Cross-sectional dependence concerns the
case in which units in the panel are influenced by unobserved global factors
and also naturally arises in our multi-country environment in which the
economies are intrinsically interconnected, due to the potential occurrence
of unobserved common global shocks.

Both issues can usefully be addressed by employing the common correlated
effect mean group estimator (CCEMG) originally developed by Pesaran (2006).
The CCEMG estimator approximates the impact of unobserved common factors
driving cross-unit dependencies through cross-sectional averages of the
variables. This empirical strategy proves to be particularly effective when
both the cross-sectional and time-series dimensions of the panel are large,
ensuring reliable estimation of common effects. Chudik and Pesaran (2015)
extend the CCEMG methodology to include dynamics. This extension
incorporates lags of both the dependent variable and the cross-sectional
averages, enabling the model to capture the persistence of common shocks and
account for serial correlation in the unobserved factors. The dynamic
framework provides bias-corrected estimators and supports robust inference,
even under the nonstationarity of observed variables or latent factors.

Based on the above considerations, we specify our empirical model as follows:%
\begin{equation}
s_{it}=\phi s_{i,t-1}+\rho b_{i,t-1}+\beta _{0}+\beta _{y}\tilde{y}%
_{i,t}+\beta _{g}\tilde{g}_{i,t}+\beta _{a}a_{i,t}+\psi
D_{t}^{GFC}+\sum_{m=0}^{3}\delta _{m}^{^{\prime }}w_{t-m}+\epsilon _{it},
\label{DCCEMG}
\end{equation}%
where $i$ indexes the cross-sectional units, $\tilde{y}_{i,t}$ and $\tilde{g}%
_{i,t}$ denote measures of temporary output and temporary spending,
respectively, $a_{i,t}$ is the current account balance, $D_{t}^{GFC}$ is a
time dummy variable equal to 1 for the period from the onset of the 2008
Global Financial Crisis onward and 0 otherwise, $w_{t}$ is the vector of
cross-sectional averages of the regressors,\footnote{%
Following Chudik and Pesaran (2015), the number of lags of the cross-sectional averages
entering our empirical specification---equal to $3$---is chosen approximately
as the cube root of the time-series length---equal to $32$.}%
  $\epsilon _{it}$ is the idiosyncratic
error term, and $\left( \phi ,\rho ,\beta _{0},\beta _{y},\beta _{g},\beta
_{a},\psi ,\delta _{m}^{\prime }\right) $ are regression coefficients.

\section{Data}

General government budget annual data on primary balances and end-of-period
gross debt for a panel of 52 industrial and emerging market economies from
1990 to 2022 are collected from the IMF's \textit{Public Finances in Modern
History} and \textit{Government Finance Statistics Yearbook}. Both variables
are scaled by the GDP to obtain the $s_{it}$ and $b_{i,t}$ time series, with
GDP data obtained from the OECD and the World Bank's \textit{National
Accounts}. Figure 1 shows the average time series of the primary balance and
government debt ratios for the entire panel and for subgroups defined by
debt levels---split at the sample median---and by country type---industrial
versus emerging.

The average debt-to-GDP ratio over the full panel shows a decreasing pattern up
to the Global Financial Crisis erupted in 2007 and, thereafter, exhibits a
massive surge---reaching two peaks, $65.2\%$ in 2016 and $74.8\%$ in 2020
in the middle of the global pandemic crisis. The dynamics differs when
splitting the panel into high-and low-debt countries and into industrial and
emerging countries. In particular, high-debt and industrial groups of
countries do not show a clear reduction in the debt-to-GDP ratio in the
pre-Global Financial Crisis period.

For the average primary balance-to-GDP ratio over the full panel, negative
peaks are dominated by the Great Recession and the Covid-19 periods ($-2.4\%$
in 2009 and $-5.1\%$ in 2020). Positive increases are visible from 1993 to
2000, during the pre- and post-Great Recession periods---specifically from
2003 to 2006 and from 2009 to 2018, and in the post-Covid-19 period. Phases
of fiscal retrenchment are detectable also for high-debt economies, and are
more pronounced for industrial countries relative to emerging
countries---especially with reference to the post-Global Financial Crisis
period.

Temporary output $\tilde{y}_{i,t}$ and temporary spending $\tilde{g}_{i,t}$
are obtained by detrending the real GDP and the real government consumption
expenditure, drawn from the OECD and the World Bank's \textit{National
Accounts, }using the Hodrick-Prescott filter with the smoothing parameter
set at 100. The resulting gaps, expressed in percentage terms, are displayed
in Figure 2.

For the output gap averaged over the whole panel, the major negative peaks
are visible in the eraly 1990s ($-1.4\%$ in 1993), in the aftermath of the
September 11 terrorist attacks ($-1.5\%$ in 2002), over the Great Recession (%
$-1.5\%$ in 2009), in the Double Dip Recession, ($-1.0\%$ in 2012-2013),
and during the pandemic crisis ($-4.2\%$ in 2020). Such peaks appear to be
quite homogeneous in terms of magnitude both across high- and low-debt
country groups and across industrial and emerging country groups. The only
exception is the Double Dip Recession of 2012-2013, which was remarkable
more severe for high-debt and industrial groups of countries, due to the
fiscal austerity measures implemented to rule out potential unsustainability
problems.

For the spending gap averaged over the whole panel, the major positive peaks
occurred mainly to counter recessions: $2.7\%$ in 1993, $2.1\%$ in 2009, and
$1.5\%$ in 2021. To offset negative peaks in the output gap, low-debt
countries reacted more than high-debt countries in 1993 and 2021, and
industrial countries reacted more than emerging countries in 2009 and 2021.

Finally, the current account balance scaled by the GDP, $a_{i,t}$, is
obtained from the IMF's \textit{International Financial Statistics} and
\textit{Balance of Payments Statistics Yearbook}.

\section{Empirical Results}

From the diagnostic tests provided in Table 1, the null hypotheses of slope
homogeneity, cross-sectional independence, and nonstationarity of the
variables are rejected. In the presence of slope heterogeneity and
cross-sectional dependence, standard panel estimation techniques typically
yield biased and inconsistent estimates, leading to invalid inferences. In addition, when cross-sectional
dependence issues arise, standard errors are often underestimated,
producing overstated significance of coefficients (Driscoll and Kraay, 1998).
This justifies our use of the dynamic common correlated effect mean group estimator
à la Chudik and Pesaran (2015).

Table 2 shows estimates of equation (\ref{DCCEMG}). Regressions 1-2 give the
results for the whole panel. Regression 1 uses only output and spending gaps
as the non-debt determinants of the primary surplus-to-GDP ratio. Regression 2
adds the current account balance. In both models, the regression coefficient
on the outstanding debt-to-GDP ratio is positive and highly significant, in
favor of the sustainability hypothesis. The signs and significance of the
regression coefficients on output and spending gaps are consistent with the
tax-smoothing hypothesis: temporary output enters positively and temporary
spending negatively, indicating the occurrence procyclical surpluses during
expansions and countercyclical deficits during downturns. There is no
significantly negative shift in the policy function following the 2008
Global Financial Crisis, which therefore appear to have acted as a temporary
shock---not systematically affecting the budget surplus policy. For
Regression 2, the current account enters positively, lending support to a
\textquotedblleft twin deficits\textquotedblright \ relationship between
external and fiscal imbalances.

The estimated coefficients are quantitatively and economically meaningful.
For the extended empirical model (Regression 2), the $\rho $- and $\phi $%
-values are $0.033$ (with standard error $=0.008$) and $0.358$ (with
standard error $=0.043$), respectively. This means that an increase in the
debt-to-GDP ratio by $10$ percentage points raises the long-run
primary-surplus-to-GDP ratio by $\left( 10\times 0.033\right) /(1-0.358)\approx
0.51$ percentage points. This policy reaction to debt accumulation provides
strong evidence of intertemporal fiscal solvency. The $\beta _{y}$- and $%
\beta _{g}$-values are $0.219$ (with standard error $=0.041$) and $-0.150$
(with standard error $=0.041$), indicating that a recession implying a fall
in the output gap by $1$ percentage point and a temporary fiscal action
implying a rise in the spending gap by $1$ percentage point \textit{per se }%
engender a cumulative increase in the primary deficit-GDP ratio (or a
cumulative decrease the primary surplus-to-GDP ratio) by $0.219/\left(
1-0.358\right) \approx 0.34$ and $0.150/\left( 1-0.358\right) \approx 0.23$
percentage points, respectively, in order to smooth taxes efficiently over
time. The $\beta _{a}$-value is $-0.085$ (with standard error $=0.040$),
indicating that an increase in the current account deficit-GDP ratio by $1$
percentage point is associated with a cumulative increase in the primary
deficit-GDP ratio by $0.085/\left( 1-0.358\right) \approx 0.13$ percentage
points.

Splitting the panel into high- and low-debt countries yields Regressions
3-6. The conditional response of primary surpluses to debt remains positive
and highly significant across the two groups. From Regressions 4 and 6, in
particular, the permanent upward adjustment in the primary surplus-to-GDP ratio
in response to a $10$-percentage-point increase in the debt-to-GDP ratio for
the high-debt group is equal to $0.47$ percentage points---$28$ percent\
lower compared to the low-debt group (equal to $0.65$ percentage points),
but still sufficient for the long-term sustainability of fiscal policy.
Observe further that tax-smoothing objectives appear to be more easily
pursued in low-debt countries, to the extent that they have a higher fiscal space.

It is worth emphasizing that, for Regressions 3-4, the 2008 Global Financial
Crisis has acted as a permanent negative shock on the level of primary
balances in high-debt countries. For Regression 3, in particular, the $\psi $
coefficient on the dummy variable $D_{t}^{GFC}$ is $-0.563$ (with standard
error $=0.314$). This means that the crisis \textit{per se} has implied a
permanent decrease in the primary surplus-to-GDP ratio by $0.563$ percentage
point. Nevertheless, it has not significantly affected the stance of
budgetary policy in terms of positive responsiveness of primary balances to
debt---which is the key criterion that matters for ensuring intertemporal
public solvency.

Splitting the panel into industrial and emerging countries yields
Regressions 7-10. The $\rho $-value continues to be positive across the two
groups, and significant at conventional levels in Regressions 7, 8, and 10.
Now the difference in the fiscal behavior is more marked, both in terms of
conditional response of primary balances to debt and in terms of tax
smoothing. From Regressions 8 and 10, in particular, the long-run budgetary
adjustment to a $10$-percentage-point increase in the debt-to-GDP ratio for
emerging countries is equal to $0.27$ percentage points---$52$\ percent
lower compared to industrial countries (equal to $0.60$ percentage points),
but again compatible with long-term sustainability.

Taking stock, the foregoing findings provide sound empirical support to the
view that a primary-balance feedback rule incorporating tax-smoothing
objectives and responding to changes in outstanding debt provides a reliable
characterization of the behavior of fiscal policymakers. Despite
heterogeneity in the magnitude of responses, the fundamental mechanism of
fiscal correction to rising debt levels is robust across country groups.
Therefore, by virtue of the fiscal requirements for public solvency
characterized in Section 2, our empirical results strongly suggest that a
large number of governments in advanced and emerging economies do not resort
to Ponzi's schemes and satisfy the intertemporal budget constraint through
the application of a Ricardian\ budgetary policy design.

\section{Conclusion}

Do fiscal authorities conduct budgetary policy without creating the potential for government bankruptcy? In the aftermath of the global financial and pandemic crises—which sharply raised debt levels—this is a pressing question for public finance analysis. Using a balanced panel of $52$ industrial and emerging market economies from 1990 onward, this paper investigates the presence and strength of fiscal correction mechanisms. It employs econometric techniques accounting for cross-sectional dependence and heterogeneity in fiscal behavior to identify the scope for a primary balance rule consistent with tax-smoothing motives and responsiveness to debt dynamics. Results show that, on average, a $10$-percentage-point increase in the debt-to-GDP ratio leads to a long-run $0.5$-percentage-point increase in the primary surplus-to-GDP ratio, controlling for cyclical output, temporary spending, and the current account. The estimated fiscal response remains significantly positive across high- and low-debt countries, and across advanced and emerging economies. While the 2008 Global Financial Crisis had a permanent negative effect on the level of primary balances in high-debt countries, it did not significantly alter fiscal policy conduct regarding the responsiveness of primary balances to debt—the central criterion for long-term public finance sustainability. Thus, the evidence supports the view that many governments maintain Ricardian fiscal behavior and satisfy the intertemporal budget constraint, avoiding Ponzi-type financing.

\newpage

\section*{References}

\begin{description}
\item[\hspace{-1cm}] Barro, R. J. (1979), \textquotedblleft On the
Determination of Public Debt,\textquotedblright \ \textit{Journal of
Political Economy} 87, 940-971.\vspace{-0.11in}

\item[\hspace{-1cm}] Barro, R. J. (1986), \textquotedblleft U.S. Deficits
since World War I,\textquotedblright \ \textit{Scandinavian Journal of
Economics} 88, 195-222.\vspace{-0.11in}

\item[\hspace{-1cm}] {Benhabib, J., S. Schmitt-Groh\'{e}\ and M. Uribe
(2001), \textquotedblleft Monetary Policy and Multiple
Equilibria,\textquotedblright\ \textit{American Economic Review }}91,
167-186.\vspace{-0.11in}

\item[\hspace{-1cm}] Blomquist, J. and J. Westerlund (2013), Testing Slope
Homogeneity in Large Panels with Serial Correlation,\textquotedblright\
\textit{Economics Letters}, 121, 374-378.\vspace{-0.11in}

\item[\hspace{-1cm}] Bohn, H. (1995), \textquotedblleft The Sustainability
of Budget Deficits in a Stochastic Economy,\textquotedblright\ \textit{%
Journal of Money, Credit and Banking} 27, 257-271.\vspace{-0.11in}

\item[\hspace{-1cm}] Bohn, H. (1998), \textquotedblleft The Behavior of U.S.
Public Debt and Deficits,\textquotedblright\ \textit{Quarterly Journal of
Economics} 113, 949-963.\vspace{-0.11in}

\item[\hspace{-1cm}] Bohn, H. (2007), \textquotedblleft Are Stationarity and Cointegration
Restrictions Really Necessary for the Intertemporal Budget
Constraint?,\textquotedblright \textit{\ Journal of Monetary Economics} 54,
1837-1847.\vspace{-0.11in}

\item[\hspace{-1cm}] Bohn, H. (2008), \textquotedblleft The Sustainability
of Fiscal Policy in the United States,\textquotedblright\ in R. Neck and J.
Sturm (Eds.), \textit{Sustainability of Public Debt}, Cambridge, MA: MIT
Press.\vspace{-0.11in}

\item[\hspace{-1cm}] Canofari, P., G. Marini and A. Piergallini (2020),
\textquotedblleft Financial Crisis and Sustainability of US Fiscal Deficit:
Indicators or Tests?,\textquotedblright\ \textit{Journal of Policy Modeling}
42, 102-204.\vspace{-0.11in}

\item[\hspace{-1cm}] Camarero, M., J. L. Carrion-i-Silvestre and C. Tamarit
(2015), \textquotedblleft The Relationship between Debt Level and Fiscal
Sustainability in Organization for Economic Cooperation and Development
Countries,\textquotedblright\ \textit{Economic Inquiry} 53, 129-149.\vspace{%
-0.11in}

\item[\hspace{-1cm}] Canzoneri, M., R. Cumby and B. Diba (2001),
\textquotedblleft Is the Price Level Determined by the Needs of Fiscal
Solvency?,\textquotedblright\ \textit{American Economic Review} 91,
1221-1238.\vspace{-0.11in}

\item[\hspace{-1cm}] Canzoneri, M. B., R. Cumby and B. Diba (2011),
\textquotedblleft The Interaction between Monetary and Fiscal
Policy,\textquotedblright\ in B. Friedman and M. Woodford (Eds.), \textit{%
Handbook of Monetary Economics} 1, Amsterdam and Boston: North-Holland
Elsevier.\vspace{-0.11in}

\item[\hspace{-1cm}] Chudik, A. and M. H. Pesaran (2015), \textquotedblleft
Common Correlated Effects Estimation of Heterogeneous Dynamic Panel Data
Models with Weakly Exogenous Regressors,\textquotedblright\ \textit{Journal
of Econometrics} 188, 393-420.\vspace{-0.11in}

\item[\hspace{-1cm}] D'Erasmo, P., E. G. Mendoza and J. Zhang (2016),
\textquotedblleft What Is a Sustainable Public Debt?,\textquotedblright\ in
J. B. Taylor and H. Uhlig (Eds.),\textit{\ Handbook of Macroeconomics} 2,
Amsterdam: Elsevier Press.\vspace{-0.11in}

\item[\hspace{-1cm}] Driscoll, J. C. and A. C. Kraay (1998), \textquotedblleft
 Consistent Covariance Matrix Estimation with Spatially Dependent
 Panel Data,\textquotedblright\ \textit{Review of Economics and Statistics} 80, 549–560.\vspace{-0.11in%
}

\item[\hspace{-1cm}] Fan, J., Y. Liao and J. Yao (2015), \textquotedblleft
Power Enhancement in High-Dimensional Cross Sectional
Tests,\textquotedblright\ \textit{Econometrica} 83, 1497-1541.\vspace{-0.11in%
}

\item[\hspace{-1cm}] McCallum, B. (1984), \textquotedblleft Are
Bond-Financed Deficits Inflationary? A Ricardian
Analysis,\textquotedblright\ \textit{Journal of Political Economy} 92,
125-135.\vspace{-0.11in}

\item[\hspace{-1cm}] Mendoza, E. G. and J. D. Ostry (2008),
\textquotedblleft International Evidence on Fiscal Solvency: Is Fiscal
Policy \textquotedblleft Responsible\textquotedblright ?,\textquotedblright\
\textit{Journal of Monetary Economics} 55, 1081-1093\vspace{-0.11in}

\item[\hspace{-1cm}] Pesaran, M. H. (2004), \textquotedblleft General Diagnostic Tests for
Cross Section Dependence in Panels,\textquotedblright\ available at \textit{%
SSRN 572504}.\vspace{-0.11in}

\item[\hspace{-1cm}] Pesaran, M. H. (2006), \textquotedblleft Estimation and
Inference in Large Heterogeneous Panels with a Multifactor Error
Structure,\textquotedblright\ \textit{Econometrica} 74, 967-1012.\vspace{%
-0.11in}

\item[\hspace{-1cm}] Pesaran, M. H. (2007), \textquotedblleft A Simple Panel
Unit Root Test in the Presence of Cross Section
Dependence,\textquotedblright\ \textit{Journal of Applied Econometrics} 22,
265-312.\vspace{-0.11in}

\item[\hspace{-1cm}] Pesaran, M. H. and T. Yamagata (2008),
\textquotedblleft Testing Slope Homogeneity in Large
Panels,\textquotedblright\ \textit{Journal of Econometrics} 142, 50-93.%
\vspace{-0.11in}
\end{description}

\newpage

\section*{Figures and Tables}

\begin{figure}[H]
\caption{Primary balance and public debt as percent of GDP, 1990--2022}
\begin{center}
\includegraphics[scale=1.0]{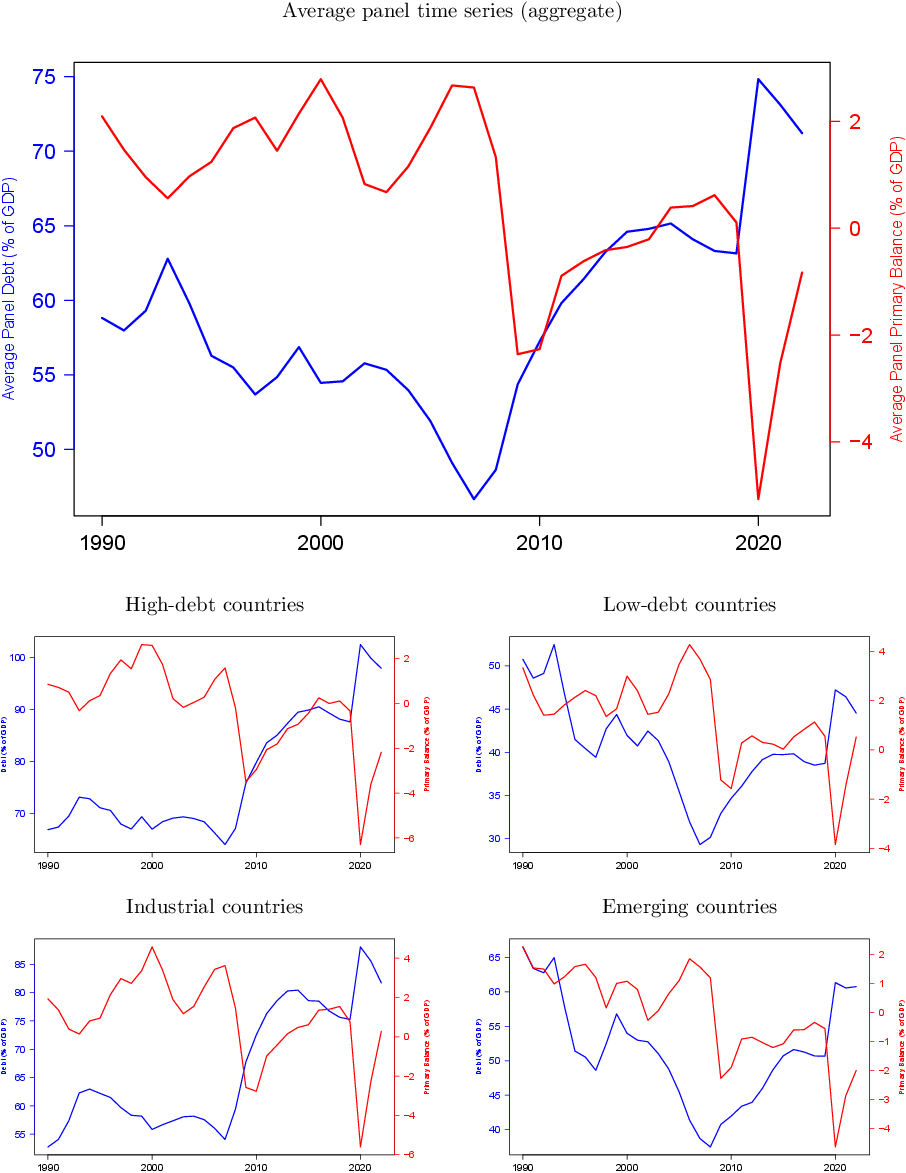}
\end{center}
\end{figure}
\newpage
\par
\begin{minipage}{1\textwidth}
\onehalfspacing
    \footnotesize
    \textit{Notes:} The panel of \textit{high-debt countries} consists of countries with median debt-to-GDP ratios above the
    median of all countries in the sample and includes: Austria, Belgium, Brazil, Canada, China, Croatia,
    Egypt, Finland, France, Germany, Greece, Hungary, Iceland, India, Israel, Italy, Japan,
    Jordan, Malaysia, Morocco, Portugal, South Africa, Spain, Ukraine, United Kingdom, United States.
    The panel of \textit{low-debt countries} consists of countries with median debt-to-GDP ratios at or below the
    median of all countries in the sample and includes: Australia, Bulgaria, Chile, Colombia,
    Cote d'Ivoire, Denmark, Indonesia, Ireland, Korea Rep., Luxembourg, Mexico,
    Netherlands, New Zealand, Nigeria, Norway, Panama, Paraguay, Peru, Philippines,
    Poland, Romania, Russian Federation, Sweden, Switzerland, Thailand, Uruguay.
    The panel of \textit{industrial countries} includes:  Australia, Austria, Belgium, Canada, China,
    Denmark, Finland, France, Germany, Greece, Iceland, Ireland, Israel, Italy, Japan, Korea Rep., Luxembourg,
    Netherlands, New Zealand, Norway, Poland, Spain, Sweden, Switzerland, United Kingdom, United States.
    The panel of \textit{emerging countries} includes: Brazil, Bulgaria, Chile, China, Colombia,
    Cote d'Ivoire, Croatia, Egypt, Hungary, India,  Indonesia, Jordan, Malaysia, Mexico, Morocco, Nigeria,
    Panama, Paraguay, Peru, Philippines, Portugal, Romania, Russian Federation, South Africa, Thailand, Ukraine.
    \end{minipage}

\begin{figure}[H]
\caption{Temporary output and temporary spending, 1990--2022}
\begin{center}
\includegraphics[scale=1.0]{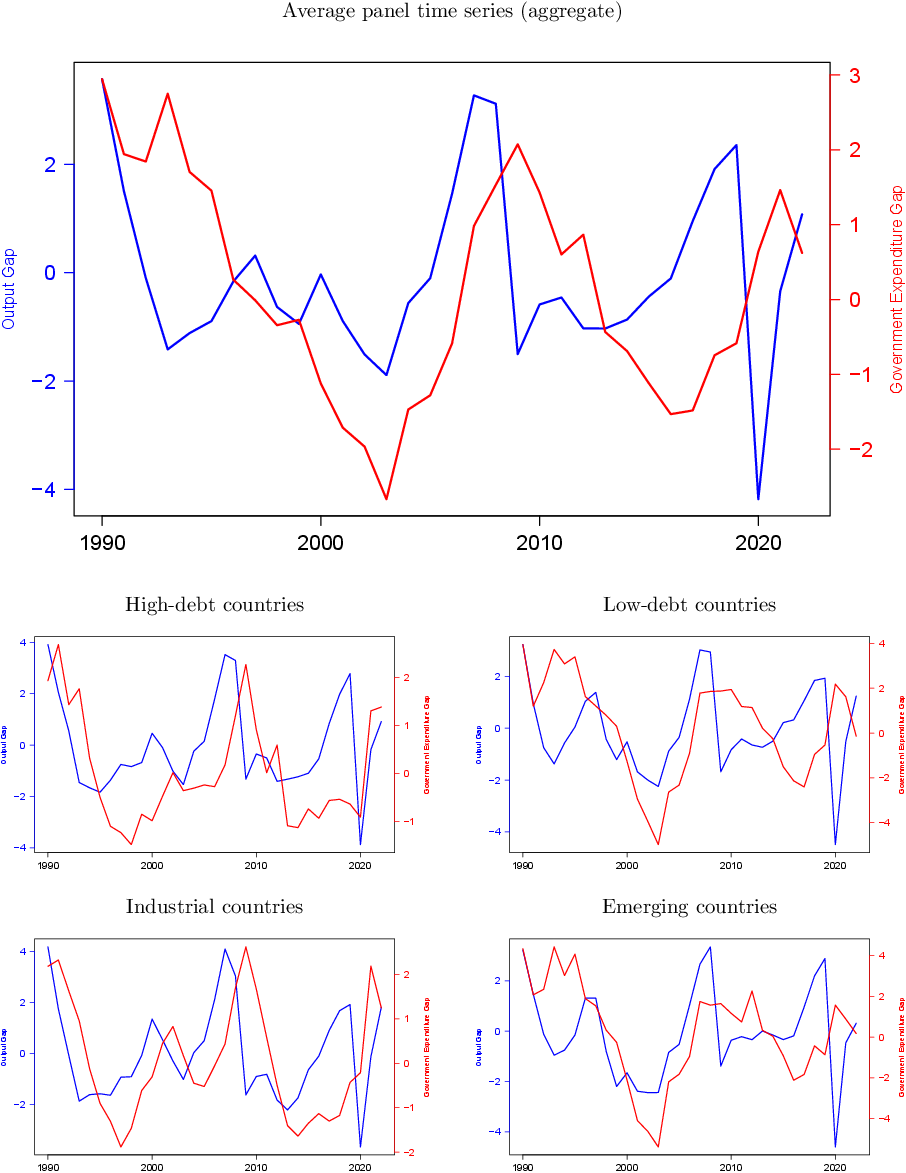}
\end{center}
\par
\vspace{0.5em}
\begin{minipage}{1\textwidth}
\onehalfspacing
    \footnotesize
    \textit{Notes:} Percentage differences from the trend computed using the Hodrick-Prescott
    filter with the smoothing parameter set at 100. Countries as in Notes to Figure 1.
    \end{minipage}
\end{figure}

\begin{sidewaystable}
\onehalfspacing
\centering
\caption{Summary and diagnostic tests, 1990--2022}
\label{tab:summary_diag}
\footnotesize
\def\sym#1{\ifmmode^{#1}\else\(^{#1}\)\fi}
\begin{tabular}{@{\,}l|@{\,}r@{\,}r@{\,}r@{\,}r@{\,}r|rrrr}
\toprule
& \multicolumn{5}{c|}{\textbf{Summary statistics}} & \multicolumn{4}{c}{\textbf{Diagnostic statistics}} \\
\cmidrule(lr){2-6} \cmidrule(lr){7-10}
& Mean & Median & SD & Min & Max & CD & CD+ & CADF & CIPS \\
\midrule
Primary balance-to-GDP ratio & -0.5097 & -0.3998 & 0.0338 & -0.1915 & 0.1421 & 0.67 (0.500) & 2776.63 (0.000) & 208.104 (0.0000) & -6.3310 (0.0000) \\
debt-to-GDP ratio & 58.9890 & 52.5500 & 36.3432 & 3.9000 & 260.1000 & -0.04 (0.970) & 3389.32 (0.000) & 166.8103 (0.0001) & -2.6080 (0.0046) \\
Output gap & -0.0385 & -0.3150 & 3.4426 & -18.1922 & 14.5874 & 0.80 (0.422) & 3066.38 (0.000) & 708.1363 (0.0000) & -19.7499 (0.0000) \\
Spending gap & 0.1548 & -0.0530 & 7.2116 & -83.7672 & 87.3504 & -2.84 (0.004) & 1863.48 (0.000) & 524.840 (0.0000) & -16.0996 (0.0000) \\
Current account-to-GDP ratio & -0.0469 & -0.6000 & 5.6287 & -23.9000 & 55.4000 & -0.32 (0.752) & 2148.49 (0.000) & 225.7231 (0.0000) & -5.9652 (0.0000) \\
\bottomrule
\end{tabular}

\vspace{0.5em}
\begin{minipage}{1\textwidth}
\footnotesize
\textit{Notes:} The Blomquist–Westerlund slope homogeneity test is 46.073 (0.0000) (Blomquist and Westerlund, 2013).
SD = standard deviation; CD = Pesaran test of cross-sectional dependence (Pesaran, 2004); CD+ = Fan–Liao–Yao power-enhanced test of cross-sectional dependence (Fan, Liao and Yao, 2015); CADF = Cross-Sectionally Augmented Dickey–Fuller test (Pesaran, 2007); CIPS = Cross-Sectionally Im–Pesaran–Shin test (Pesaran, 2007).
P-values are in parentheses. The panel includes: Australia, Austria, Belgium, Brazil, Bulgaria, Canada, Chile, China, Colombia, Côte d'Ivoire, Croatia, Denmark, Egypt, Indonesia, Ireland, Korea Rep., Finland, France, Germany, Greece, Hungary, Iceland, India, Israel, Italy, Japan, Jordan, Luxembourg, Malaysia, Mexico, Morocco, Netherlands, New Zealand, Nigeria, Norway, Panama, Paraguay, Peru, Philippines, Poland, Portugal, Romania, Russian Federation, South Africa, Spain, Sweden, Switzerland, Thailand, Ukraine, United Kingdom, United States, Uruguay.
Since the CD and CD+ tests yield conflicting results, we tend to rely on the CD+ outcome due to its improved power and consistency in panels with a large time-series dimension (Fan, Liao and Yao, 2015).
\end{minipage}
\end{sidewaystable}

\begin{sidewaystable}
\onehalfspacing
\centering
\caption{Determinants of the primary balance-to-GDP ratio, 1990--2022}
\label{tab:determinants_pb}
\footnotesize
\def\sym#1{\ifmmode^{#1}\else\(^{#1}\)\fi}
\begin{tabular}{l|cc|cc|cc|cc|cc}
\toprule
& \multicolumn{2}{c}{Aggregate panel}
& \multicolumn{2}{c}{High-debt countries}
& \multicolumn{2}{c}{Low-debt countries}
& \multicolumn{2}{c}{Industrial countries}
& \multicolumn{2}{c}{Emerging countries} \\
& (1) & (2) & (3) & (4) & (5) & (6) & (7) & (8) & (9) & (10) \\
\midrule
Lagged primary balance-to-GDP ratio
& 0.427\sym{***} & 0.358\sym{***} & 0.357\sym{***} & 0.308\sym{***} & 0.446\sym{***} & -0.377\sym{***} & 0.400\sym{***} & 0.336\sym{***} & 0.383\sym{***} & 0.323\sym{***} \\
& (0.0406) & (0.0427) & (0.0658) & (0.0670) & (0.0494) & (0.0513) & (0.0511) & (0.0507) & (0.0589) & (0.0627) \\
Lagged debt-to-GDP ratio
& 0.0234\sym{**} & 0.0331\sym{***} & 0.0300\sym{**} & 0.0344\sym{**} & 0.0308\sym{**} & 0.0403\sym{***} & 0.0222\sym{**} & 0.0371\sym{**} & 0.0158 & 0.0182\sym{**} \\
& (0.00771) & (0.00809) & (0.0112) & (0.0108) & (0.0105) & (0.0120) & (0.0109) & (0.0114) & (0.0098) & (0.0092) \\
Constant
& -1.358\sym{**} & -1.905\sym{**} & -1.893\sym{**} & -2.233\sym{**} & -1.226\sym{*} & -1.561\sym{**} & -1.300\sym{*} & -2.393\sym{**} & -1.015 & -0.686 \\
& (0.565) & (0.588) & (0.726) & (0.703) & (0.662) & (0.715) & (0.742) & (0.769) & (0.649) & (0.582) \\
Output gap
& 0.202\sym{***} & 0.219\sym{***} & 0.202\sym{***} & 0.188\sym{**} & 0.237\sym{***} & 0.257\sym{***} & 0.310\sym{***} & 0.290\sym{***} & 0.0755\sym{**} & 0.138\sym{**} \\
& (0.0455) & (0.0411) & (0.0535) & (0.0569) & (0.0705) & (0.0610) & (0.0726) & (0.0672) & (0.0380) & (0.0498) \\
Spending gap
& -0.157\sym{***} & -0.150\sym{***} & -0.142\sym{**} & -0.128\sym{**} & -0.154\sym{***} & -0.141\sym{**} & -0.233\sym{***} & -0.201\sym{**} & -0.0910\sym{**} & -0.0919\sym{**} \\
& (0.0363) & (0.0384) & (0.0660) & (0.0639) & (0.0391) & (0.0456) & (0.0648) & (0.0646) & (0.0362) & (0.0343) \\
2008 Global Financial Crisis dummy
& -0.295 & -0.357 & -0.563\sym{*} & -0.600\sym{*} & -0.335 & -0.434 & -0.113 & -0.0203 & -0.298 & -0.588 \\
& (0.320) & (0.321) & (0.314) & (0.321) & (0.488) & (0.508) & (0.321) & (0.344) & (0.513) & (0.558) \\
Current account-to-GDP ratio
& -- & 0.0851\sym{**} & -- & -0.00294 & -- & 0.157\sym{***} & -- & 0.0870\sym{*} & -- & 0.0813 \\
& -- & (0.0391) & -- & (0.0519) & -- & (0.0470) & -- & (0.0517) & -- & (0.0577) \\
\bottomrule
\end{tabular}

\vspace{0.5em}
\begin{minipage}{1\textwidth}
\footnotesize
\textit{Notes:} Estimates of equation (\ref{DCCEMG}); standard errors are in parentheses;
\sym{*}, \sym{**}, \sym{***} indicate p-values at the 10\%, 5\%, and 1\% levels, respectively.
The panel of \textit{high-debt countries} consists of countries with median debt-to-GDP ratios above the
    median of all countries in the sample and includes: Austria, Belgium, Brazil, Canada, China, Croatia,
    Egypt, Finland, France, Germany, Greece, Hungary, Iceland, India, Israel, Italy, Japan,
    Jordan, Malaysia, Morocco, Portugal, South Africa, Spain, Ukraine, United Kingdom, United States.
The panel of \textit{low-debt countries} consists of countries with median debt-to-GDP ratios at or below the
    median of all countries in the sample and includes: Australia, Bulgaria, Chile, Colombia,
    Cote d'Ivoire, Denmark, Indonesia, Ireland, Korea Rep., Luxembourg, Mexico,
    Netherlands, New Zealand, Nigeria, Norway, Panama, Paraguay, Peru, Philippines,
    Poland, Romania, Russian Federation, Sweden, Switzerland, Thailand, Uruguay.
The panel of \textit{industrial countries} includes:  Australia, Austria, Belgium, Canada, China,
    Denmark, Finland, France, Germany, Greece, Iceland, Ireland, Israel, Italy, Japan, Korea Rep., Luxembourg,
    Netherlands, New Zealand, Norway, Poland, Spain, Sweden, Switzerland, United Kingdom, United States.
The panel of \textit{emerging countries} includes: Brazil, Bulgaria, Chile, China, Colombia,
    Cote d'Ivoire, Croatia, Egypt, Hungary, India,  Indonesia, Jordan, Malaysia, Mexico, Morocco, Nigeria,
    Panama, Paraguay, Peru, Philippines, Portugal, Romania, Russian Federation, South Africa, Thailand, Ukraine.
\end{minipage}
\end{sidewaystable}

\end{doublespace}

\end{document}